# Conservative and skew-symmetric forms of the incompressible Navier-Stokes equations in sigma-coordinates

Jaeyoung Jung[1,*] & Marco Giometto[1]

[1]Department of Civil Engineering and Engineering Mechanics, Columbia University,
New York, NY 10027

**Abstract.** This study derives conservative and skew-symmetric formulations of the incompressible flow equations in a terrain-following $\sigma$-coordinate system that preserve key structural properties of the Cartesian formulation. Unlike conventional formulations based on the direct application of the $\sigma$-transformation to Cartesian equations, in which metric-induced terms disrupt the intrinsic structure of the governing equations, the proposed formulations are designed to avoid these structural inconsistencies. A conservative form is derived in a manner consistent with general conservation laws, and its modified eigenstructure is analyzed relative to the Cartesian counterpart. A skew-symmetric formulation is then derived by introducing a new set of variables, yielding a form that is energy-conserving for the Euler equations and energy-bounded for the Navier–Stokes equations. Finally, we discuss characteristic-based boundary conditions to ensure energy boundedness of the system.

# 1. Introduction

When the boundary geometry becomes highly complex, such as in atmospheric boundary-layer flows over complex terrain and air-water interfaces under strong wind conditions, conventional approaches typically rely on either body-conforming grids or immersed boundary methods to represent boundaries [1,2]. While body-conforming grids provide high geometric fidelity, they require substantial meshing effort and computational cost, and may not be feasible for highly complex or evolving geometries [1]. In contrast, Cartesian-grid-based approaches, including immersed boundary methods, avoid complex meshing but introduce significant complications in the treatment of boundaries [2]. To accurately capture the influence of complex boundaries on the flow field, the $\sigma$-transformation, which rescales the vertical coordinate according to the terrain height, has been widely adopted [3]. Despite its practical convenience, the $\sigma$-coordinate approach breaks the local orthogonality between horizontal and vertical directions. As a result, numerous metric-induced terms arise, leading to a substantial increase in the algebraic and structural complexity of the governing equations [3,4].

Nevertheless, due to its effectiveness in handling irregular boundaries, the $\sigma$-coordinate system has been extensively used in atmospheric and oceanic modeling efforts [5–7]. In most existing schemes, the conservative or advective forms defined in Cartesian coordinates are transformed in a straightforward manner into the $\sigma$-coordinate system (e.g., [4]). Owing to the presence of metric-induced terms, the transformed equations do not, in general, preserve a strictly conservative or skew-symmetric structure. While this procedure allows existing derivative operators to be reused with minor modifications, it does not guarantee that the fundamental structural properties of the original equations—such as the shock-capturing capability of conservative formulations or the energy-conserving (or energy-bounded) property of skew-symmetric forms—are preserved under the coordinate transformation.

In the present study, we demonstrate that both conservative and skew-symmetric formulations can be rigorously constructed within the $\sigma$-coordinate framework. For the conservative formulation, the altered eigen-structure induced by the $\sigma$-transformation is derived and compared with that of the Cartesian system. In addition, the reformulation strategies proposed in [8–11] are extended to the $\sigma$-coordinate framework to construct a skew-symmetric formulation. The energy method [12] confirms that the resulting system is energy-conserving in the inviscid (Euler) case and energy-bounded in the viscous (Navier–Stokes) case.

## 2. Conservative form of conservation laws in the sigma-coordinate transform

The derivation begins with the system of conservation laws in the Cartesian-coordinate system:

$$\frac{\partial \mathbf{U}}{\partial t} + \frac{\partial \mathbf{F}}{\partial x} + \frac{\partial \mathbf{G}}{\partial z} = \mathbf{S} \tag{1}$$

where $\mathbf{U}$ is the vector of conservative variables, $\mathbf{F}$ and $\mathbf{G}$ are the flux vectors in the $x$- and $z$-directions, respectively, and $\mathbf{S}$ denotes a source term. The terrain-following, or sigma-coordinate, system is defined by the coordinate transformation:

$$\hat{t} = t, \hat{x} = x, \sigma = \frac{z - b(x,t)}{H - b(x,t)} \tag{2}$$

where, $b$ denotes the terrain elevation, and $H$ represents the height of the computational domain. Since the horizontal and temporal coordinates remain unchanged, the hat notation is omitted hereafter for simplicity.

The value of $z$ at the $\sigma$-layer in the physical domain is given by $z = z(x, \sigma, t) = \sigma(H - b(x,t)) + b(x,t)$. For $z \in C^1$, the following relations follow from the total differential $dz$:

$$\begin{cases} dz|_x = \frac{\partial z}{\partial t}\Big|_x + \frac{\partial \sigma}{\partial t}\Big|_{x,z}\frac{\partial z}{\partial \sigma} = 0 \Rightarrow \frac{\partial \sigma}{\partial t}\Big|_{x,z} = -\frac{z_t}{z_\sigma} = -\frac{z_t}{J}, \\ dz|_t = \frac{\partial z}{\partial x}\Big|_t + \frac{\partial \sigma}{\partial x}\Big|_{z,t}\frac{\partial z}{\partial \sigma} = 0 \Rightarrow \frac{\partial \sigma}{\partial x}\Big|_{z,t} = -\frac{z_x}{z_\sigma} = -\frac{z_x}{J}, \end{cases} \tag{3}$$

where the subscript $\chi \in \{t, x, z, \sigma\}$ denotes the partial derivative with respect to $\chi$ (e.g., $z_\sigma \equiv \partial z/\partial \sigma$) and $J = z_\sigma$ is the Jacobian of the coordinate transformation. Then, the partial derivatives in Cartesian coordinates system can be rewritten in the sigma-coordinate system as follows:

$$\begin{cases} \left.\dfrac{\partial}{\partial t}\right|_z = \left.\dfrac{\partial}{\partial t}\right|_\sigma + \left.\dfrac{\partial \sigma}{\partial t}\right|_z \dfrac{\partial}{\partial \sigma} = \left.\dfrac{\partial}{\partial t}\right|_\sigma - \dfrac{z_t}{J}\dfrac{\partial}{\partial \sigma}, \\ \left.\dfrac{\partial}{\partial x}\right|_z = \left.\dfrac{\partial}{\partial x}\right|_\sigma + \left.\dfrac{\partial \sigma}{\partial x}\right|_z \dfrac{\partial}{\partial \sigma} = \left.\dfrac{\partial}{\partial x}\right|_\sigma - \dfrac{z_x}{J}\dfrac{\partial}{\partial \sigma}, \\ \left.\dfrac{\partial}{\partial z}\right|_x = \left.\dfrac{\partial \sigma}{\partial z}\right|_x \dfrac{\partial}{\partial \sigma} = \dfrac{1}{J}\dfrac{\partial}{\partial \sigma}. \end{cases} \quad (4)$$

Substituting Eq. (4) into Eq. (1) yields

$$J\dfrac{\partial \mathbf{U}}{\partial t} - z_t \dfrac{\partial \mathbf{U}}{\partial \sigma} + J\dfrac{\partial \mathbf{F}}{\partial x} - z_x \dfrac{\partial \mathbf{F}}{\partial \sigma} + \dfrac{\partial \mathbf{G}}{\partial \sigma} = J\mathbf{S}. \quad (5)$$

When $z \in C^2$, the mixed partial derivatives commute, yielding

$$\begin{cases} \dfrac{\partial^2 z}{\partial t \partial \sigma} = \dfrac{\partial^2 z}{\partial \sigma \partial t} \Rightarrow \dfrac{\partial J}{\partial t} - \dfrac{\partial (z_t)}{\partial \sigma} = 0, \\ \dfrac{\partial^2 z}{\partial x \partial \sigma} = \dfrac{\partial^2 z}{\partial \sigma \partial x} \Rightarrow \dfrac{\partial J}{\partial x} - \dfrac{\partial (z_x)}{\partial \sigma} = 0. \end{cases} \quad (6)$$

Therefore, Eq. (5) can be re-written in the conservative form of governing system as follows:

$$\dfrac{\partial \mathbf{V}}{\partial t} + \dfrac{\partial \widetilde{\mathbf{F}}}{\partial x} + \dfrac{\partial \widetilde{\mathbf{G}}}{\partial \sigma} = \widetilde{\mathbf{S}} \quad (7)$$

where $\mathbf{V} = J\mathbf{U}$, $\widetilde{\mathbf{F}} = J\mathbf{F}$, $\widetilde{\mathbf{G}} = \mathbf{G} - z_t \mathbf{U} - z_x \mathbf{F}$, and $\widetilde{\mathbf{S}} = J\mathbf{S}$.

# 3. Incompressible Euler equations in the sigma-coordinate system

## 3.1. Conservative form of the incompressible Euler equations

The two-dimensional incompressible Euler (IE) equations in the Cartesian-coordinate system, i.e., Eq. (1) with

$$\mathbf{U} = \begin{pmatrix} 0 \\ u \\ w \end{pmatrix}, \mathbf{F} = \begin{pmatrix} u \\ uu + p/\rho_o \\ wu \end{pmatrix}, \mathbf{G} = \begin{pmatrix} w \\ uw \\ ww + p/\rho_o \end{pmatrix}, \text{ and } \mathbf{S} = \begin{pmatrix} 0 \\ 0 \\ 0 \end{pmatrix}, \qquad (8)$$

can be re-written in the sigma-coordinate system as Eq. (7) with

$$\mathbf{V} = \begin{pmatrix} J \\ Ju \\ Jw \end{pmatrix}, \tilde{\mathbf{F}} = \begin{pmatrix} Ju \\ Juu + Jp/\rho_o \\ Jwu \end{pmatrix}, \tilde{\mathbf{G}} = \begin{pmatrix} w^* \\ uw^* - z_x p/\rho_o \\ ww^* + p/\rho_o \end{pmatrix}, \text{ and } \tilde{\mathbf{S}} = \begin{pmatrix} 0 \\ 0 \\ 0 \end{pmatrix} \qquad (9)$$

where $w^* = w - z_t - z_x u$ represents the Jacobian-weighted contravariant flux velocity normal to the $\sigma = const.$ surfaces. It is worth noting that $Ju$ corresponds to the Jacobian-weighted flux velocity in the $x$-direction. Since the vertical grid spacing in the $z$-direction varies with the underlying topography, the resulting horizontal flux term naturally takes the form of the volume-averaged Euler equations [13], analogous to that used in the analysis of duct flows [14]. Meanwhile, under the incompressibility assumption, the pressure exhibits a global nature. Taking the divergence of the momentum equations yields an elliptic problem for the pressure:

$$\mathcal{L}[p] = -\rho_o \mathcal{R} \qquad (10)$$

where

$$\begin{cases} \mathcal{L}[p] = \partial_{xx}(Jp) - \partial_{x\sigma}(z_x p) + \partial_{\sigma\sigma}(p), \\ \mathcal{R} = \partial_t[\partial_x(Ju) + \partial_\sigma(Jw)] + \partial_{xx}(Juu) + \partial_{x\sigma}[(w^* + Jw)u] + \partial_{\sigma\sigma}(ww^*), \end{cases} \qquad (11)$$

As Eqs. (7) and (9) are formulated in conservative form for mass and momentum, their straightforward discretization does not, in general, ensure energy conservation. In cases where an accurate representation of the flow energy structure is essential, such as in simulations of turbulent flows, the energy-conserving property (or energy boundedness) becomes a crucial requirement for numerical stability and physical fidelity. This property can be satisfied through a skew-symmetric formulation (e.g., see Proposition 3.1. in [9]), as detailed in Sect. 3.2. In contrast, the conservative form is advantageous for capturing the characteristics of the hyperbolic subsystem [15]. When incompressible flows are modeled using weakly compressible methods or pseudo-compressibility approaches, the resulting system recovers hyperbolicity. In such cases, a conservative formulation is advantageous for accurately capturing the solution structure, thereby enhancing numerical stability. The following example presents the artificial compressibility formulation of Chorin [16], which is one of the most widely used pseudo-compressibility methods.

*Example*. The artificial compressibility method for IE over the non-movable irregular terrain $z = z(x, \sigma)$ reads as Eq. (7) with

$$\mathbf{V} = \begin{pmatrix} Jp \\ Ju \\ Jw \\ J \\ z_x \end{pmatrix}, \mathbf{F} = \begin{pmatrix} \rho_o c^2 Ju \\ Juu + Jp/\rho_o \\ Jwu \\ 0 \\ 0 \end{pmatrix}, \mathbf{G} = \begin{pmatrix} \rho_o c^2 w^* \\ uw^* - z_x p/\rho_o \\ ww^* + p/\rho_o \\ 0 \\ 0 \end{pmatrix}, and\ \tilde{\mathbf{S}} = \begin{pmatrix} 0 \\ 0 \\ 0 \\ 0 \\ 0 \end{pmatrix} \quad (12)$$

where $w^* = w - z_x u$ and $c$ represents the artificial sound speed. It is worth noting that, as the artificial compressibility method is intrinsically a pseudo-time-marching approach toward a steady state, the terrain elevation is treated as invariant in pseudo-time. The last two components, corresponding to $\partial_t J = \partial_t z_x = 0$, are hence introduced solely for the purpose of eigen-structure analysis. The quasi-linear form of Eq. (7) reads:

$$\frac{\partial \mathbf{V}}{\partial t} + \frac{\partial \tilde{\mathbf{F}}}{\partial \mathbf{V}}\frac{\partial \mathbf{V}}{\partial x} + \frac{\partial \tilde{\mathbf{G}}}{\partial \mathbf{V}}\frac{\partial \mathbf{V}}{\partial \sigma} = \tilde{\mathbf{S}}, \tag{13}$$

with

$$\frac{\partial \tilde{\mathbf{F}}}{\partial \mathbf{V}} = \begin{pmatrix} 0 & \rho_o c^2 & 0 & 0 & 0 \\ 1/\rho_o & 2u & 0 & -uu & 0 \\ 0 & w & u & -uw & 0 \\ 0 & 0 & 0 & 0 & 0 \\ 0 & 0 & 0 & 0 & 0 \end{pmatrix} \tag{14}$$

and

$$\frac{\partial \tilde{\mathbf{G}}}{\partial \mathbf{V}} = \frac{1}{J}\begin{pmatrix} 0 & -\rho_o c^2 z_x & \rho_o c^2 & -\rho_o c^2 w^* & -J\rho_o c^2 u \\ -z_x/\rho_o & w - 2z_x u & u & -2uw^* + z_x p/\rho_o & -J(u^2 + p/\rho_o) \\ 1/\rho_o & -z_x w & 2w - z_x u & -2ww^* - p/\rho_o & -Juw \\ 0 & 0 & 0 & 0 & 0 \\ 0 & 0 & 0 & 0 & 0 \end{pmatrix}. \tag{15}$$

The Jacobian matrix of the $x$-directional flux term, $\partial \tilde{\mathbf{F}}/\partial \mathbf{V}$, is identical to that of the volume-averaged Euler equations in the Cartesian system (see [13]). The eigenvalues of $\partial \tilde{\mathbf{F}}/\partial \mathbf{V}$ are

$$\lambda_1^x = u + \sqrt{u^2 + c^2}, \lambda_2^x = u, \lambda_3^x = u - \sqrt{u^2 + c^2}, \lambda_4^x = \lambda_5^x = 0, \tag{16}$$

with their associated eigen-vectors

$$\begin{cases} \mathbf{r}_1^x = \left(\rho_o(c^2 + u\lambda_3^x), \sqrt{u^2 + c^2}, w, 0, 0\right)^T, \\ \mathbf{r}_2^x = (0,0,1,0,0)^T, \\ \mathbf{r}_3^x = \left(\rho_o(c^2 + u\lambda_1^x), -\sqrt{u^2 + c^2}, w, 0, 0\right)^T, \\ \mathbf{r}_4^x = (\rho_o u^2, 0, w, 1, 0)^T, \\ \mathbf{r}_5^x = (0,0,0,0,1)^T, \end{cases} \tag{17}$$

respectively. On the other hand, the Jacobian matrix of the $\sigma$-directional flux term, $\partial \tilde{\mathbf{G}}/\partial \mathbf{V}$, exhibits substantially different structure from that of the horizontal flux term, and its eigenvalues are given by

$$\begin{cases} \lambda_1^\sigma = \dfrac{1}{J}\left(w^* + \sqrt{w^{*2} + (1+z_x^2)c^2}\right), \\ \lambda_2^\sigma = \dfrac{w^*}{J}, \\ \lambda_3^\sigma = \dfrac{1}{J}\left(w^* - \sqrt{w^{*2} + (1+z_x^2)c^2}\right), \\ \lambda_4^\sigma = \lambda_5^\sigma = 0, \end{cases} \qquad (18)$$

with the associated eigen-vectors

$$\begin{cases} \mathbf{r}_1^\sigma = \left(\dfrac{\rho_o(w^* - \rho_o c^2(1+z_x^2))\lambda_3^\sigma J}{(1+z_x^2)w - \lambda_3^\sigma J}, \dfrac{(1+z_x^2)u + z_x \lambda_3^\sigma J}{(1+z_x^2)w - \lambda_3^\sigma J}, 1, 0, 0\right)^T, \\ \mathbf{r}_2^\sigma = (0, 1, z_x, 0, 0)^T, \\ \mathbf{r}_3^\sigma = \left(\dfrac{\rho_o(w^* + \rho_o c^2(1+z_x^2))\lambda_1^\sigma J}{(1+z_x^2)w - \lambda_1^\sigma J}, \dfrac{(1+z_x^2)u + z_x \lambda_1^\sigma J}{(1+z_x^2)w - \lambda_1^\sigma J}, 1, 0, 0\right)^T, \\ \mathbf{r}_4^\sigma = (p, u, w, 1, 0)^T, \\ \mathbf{r}_5^\sigma = \left(-\dfrac{(z_x p + \rho_o u w^*)J}{1+z_x^2}, \dfrac{(p/\rho_o - z_x u w^*)J}{w^*(1+z_x^2)}, \dfrac{(z_x p/\rho_o + u w^*)J}{w^*(1+z_x^2)}, 0, 1\right)^T, \end{cases} \qquad (19)$$

respectively. In Eqs. (18) and (19), the characteristic fields associated to the fourth and fifth eigen-values account for the effects of discontinuities in $J$ and $z_x$ on the solution. However, since $J$ is assumed to be sufficiently smooth (i.e., $z \in C^2$) in Sect. 2, the elementary waves associated with the fourth and fifth eigenspaces do not contribute to the solution. Consequently, only the leading $3 \times 3$ principal blocks of the Jacobian matrices (14) and (15) govern the flow dynamics. It can be shown that, in the limit $z_x \to 0$, the first three eigenstructures in the $\sigma$-direction reduce to those of the Cartesian coordinate system.

### 3.2. The skew-symmetric form of incompressible Euler equations

In the Cartesian coordinate system, the skew-symmetric form of the incompressible Euler equations is obtained by adding the continuity equation multiplied by $-0.5\mathbf{U}$ to the conservative forms of the momentum equations. Using this straightforward manipulation, Eqs. (1) and (8) can be recast into the following form:

$$\begin{cases} \dfrac{\partial J}{\partial t} + \dfrac{\partial (Ju)}{\partial x} + \dfrac{\partial (w^*)}{\partial \sigma} = 0, \\ \dfrac{\partial (Ju)}{\partial t} - \dfrac{u}{2}\dfrac{\partial J}{\partial t} + \dfrac{1}{2}\left[\dfrac{\partial (Juu)}{\partial x} + Ju\dfrac{\partial u}{\partial x}\right] + \dfrac{1}{2}\left[\dfrac{\partial (uw^*)}{\partial \sigma} + w^*\dfrac{\partial u}{\partial \sigma}\right] + \dfrac{\partial}{\partial x}\left(J\dfrac{p}{\rho_o}\right) - \dfrac{\partial}{\partial \sigma}\left(z_x\dfrac{p}{\rho_o}\right) = 0, \quad (20) \\ \dfrac{\partial (Jw)}{\partial t} - \dfrac{w}{2}\dfrac{\partial J}{\partial t} + \dfrac{1}{2}\left[\dfrac{\partial (Jwu)}{\partial x} + Ju\dfrac{\partial w}{\partial x}\right] + \dfrac{1}{2}\left[\dfrac{\partial (ww^*)}{\partial \sigma} + w^*\dfrac{\partial w}{\partial \sigma}\right] + \dfrac{\partial}{\partial \sigma}\left(\dfrac{p}{\rho_o}\right) = 0, \end{cases}$$

where $w^* = w - z_t - z_x u$. In this formulation, the convective momentum flux terms are explicitly written in skew-symmetric form and therefore admit a kinetic-energy balance through the use of summation-by-parts operators. It is important to note, however, that this property pertains to the physical kinetic-energy balance associated with the nonlinear convective terms, obtained after weighting the momentum equations by the velocity components as follows:

$$\begin{cases} \dfrac{1}{2}\left[\dfrac{\partial (Juu)}{\partial x} + Ju\dfrac{\partial u}{\partial x}\right] u + \dfrac{1}{2}\left[\dfrac{\partial (uw^*)}{\partial \sigma} + w^*\dfrac{\partial u}{\partial \sigma}\right] u = \dfrac{\partial}{\partial x}\left(Ju\dfrac{u^2}{2}\right) + \dfrac{\partial}{\partial \sigma}\left(w^*\dfrac{u^2}{2}\right), \\ \dfrac{1}{2}\left[\dfrac{\partial (Juw)}{\partial x} + Ju\dfrac{\partial w}{\partial x}\right] w + \dfrac{1}{2}\left[\dfrac{\partial (ww^*)}{\partial \sigma} + w^*\dfrac{\partial w}{\partial \sigma}\right] w = \dfrac{\partial}{\partial x}\left(Ju\dfrac{w^2}{2}\right) + \dfrac{\partial}{\partial \sigma}\left(w^*\dfrac{u^2}{2}\right). \end{cases} \quad (21)$$

In contrast, the energy method in the sense of a $L^2$-norm typically requires a formulation in which the time-derivative term closes directly in a quadratic norm, i.e., $\langle V, \partial_t V \rangle = \partial_t(0.5\|V\|^2)$, for a suitably chosen set of variables $V$. In Eq. (20), however, additional metric-dependent temporal terms (i.e., those involving $\partial_t J$) arise, thereby breaking the skew-symmetric structure of the formulation. Moreover, the quadratic norm induced by the conservative variables $(Ju, Jw)$ does not recover the physical kinetic energy. Instead of yielding the metric-weighted energy $J(u^2/2 + w^2/2)$, the energy analysis produces terms of the form $J^2/2 + (Ju)^2/2 + (Jw)^2/2$. Hence, the well-balanced treatment for pressure terms in Eq. (20) alone is insufficient to establish an energy-conserving (or energy-bounded) formulation in the $L^2$-sense without introducing an alternative skew-symmetric form in which the time-differentiated variables themselves define the energy norm.

To derive an alternative energy-conserving skew-symmetric formulation in the sigma-coordinate system, the present study introduces the transformed variables:

$$P \equiv \sqrt{J}p/\rho_o, \ U \equiv \sqrt{J}u \text{ and } W \equiv \sqrt{J}w, \tag{22}$$

whose choice is inspired by [8,10]. Substituting Eq.(22) into Eq. (20) yields:

$$\begin{cases} 0\dfrac{\partial P}{\partial t} + \dfrac{1}{2}\left[\dfrac{\partial 1 U}{\partial x} + 1\dfrac{\partial U}{\partial x}\right] + \dfrac{1}{2}\left[\dfrac{\partial}{\partial \sigma}\left(\dfrac{W}{J}\right) + \dfrac{1}{J}\dfrac{\partial W}{\partial \sigma}\right] - \dfrac{1}{2}\left[\dfrac{\partial}{\partial \sigma}\left(\dfrac{z_x U}{J}\right) + \dfrac{z_x}{J}\dfrac{\partial U}{\partial \sigma}\right] = 0, \\ \dfrac{\partial U}{\partial t} + \dfrac{1}{2}\left[\dfrac{\partial (uU)}{\partial x} + u\dfrac{\partial U}{\partial x}\right] + \dfrac{1}{2}\left[\dfrac{\partial}{\partial \sigma}\left(\dfrac{w^*}{J}U\right) + \dfrac{w^*}{J}U_\sigma\right] + \dfrac{1}{2}\left[\dfrac{\partial (1P)}{\partial x} + 1\dfrac{\partial P}{\partial x}\right] - \dfrac{1}{2}\left[\dfrac{\partial}{\partial \sigma}\left(\dfrac{z_x}{J}P\right) + \dfrac{z_x}{J}\dfrac{\partial P}{\partial \sigma}\right] = 0, \\ \dfrac{\partial W}{\partial t} + \dfrac{1}{2}\left[\dfrac{\partial (uW)}{\partial x} + u\dfrac{\partial W}{\partial x}\right] + \dfrac{1}{2}\left[\dfrac{\partial}{\partial \sigma}\left(\dfrac{w^*}{J}W\right) + \dfrac{w^*}{J}W_\sigma\right] + \dfrac{1}{2}\left[\dfrac{\partial}{\partial \sigma}\left(\dfrac{1}{J}P\right) + \dfrac{1}{J}\dfrac{\partial P}{\partial \sigma}\right] = 0. \end{cases} \tag{23}$$

A detailed derivation can be found in the Appendix. Eq. (23) can be re-written as

$$\tilde{I}\frac{\partial \mathbf{u}}{\partial t} + \frac{1}{2}[(A\mathbf{u})_x + A\mathbf{u}_x] + \frac{1}{2}[(B\mathbf{u})_\sigma + B\mathbf{u}_\sigma] = \Lambda \tilde{\mathbf{S}}, \tag{24}$$

where $\mathbf{u} = (\sqrt{J}p/\rho_o, \sqrt{J}u, \sqrt{J}w)$, $\tilde{I} = diag(0,1,1)$, $\Lambda = diag(1, J^{-1/2}, J^{-1/2})$, $\tilde{\mathbf{S}} = (0,0,0)^T$,

$$A = \begin{pmatrix} 0 & 1 & 0 \\ 1 & u & 0 \\ 0 & 0 & u \end{pmatrix}, \text{ and } B = \begin{pmatrix} 0 & -z_x/J & 1/J \\ -z_x/J & w^*/J & 0 \\ 1/J & 0 & w^*/J \end{pmatrix}. \tag{25}$$

If we apply the energy method, we have

$$\int_\Omega \mathbf{u}^T \left[\tilde{I}\frac{\partial \mathbf{u}}{\partial t} + \frac{1}{2}[(A\mathbf{u})_x + A\mathbf{u}_x] + \frac{1}{2}[(B\mathbf{u})_\sigma + B\mathbf{u}_\sigma] - \Lambda \tilde{\mathbf{S}}\right] d\Omega = \mathbf{0}, \tag{26}$$

Using the relations $\mathbf{u}^T(A\mathbf{u})_x = (\mathbf{u}^T A\mathbf{u})_x - (\mathbf{u}^T)_x A\mathbf{u}$ and $\mathbf{u}^T(B\mathbf{u})_\sigma = (\mathbf{u}^T B\mathbf{u})_\sigma - (\mathbf{u}^T)_\sigma B\mathbf{u}$ and Green's theorem, Eq. (26) becomes:

$$\frac{d\|\mathbf{u}\|_{\tilde{I}}^2}{dt} + \frac{1}{2}\oint_{\partial\Omega}(BT_x + BT_\sigma)dA = 0, \tag{27}$$

where $\|\mathbf{u}\|_{\tilde{I}}^2 \equiv \int_\Omega \mathbf{u}^T \tilde{I} \mathbf{u} d\Omega$, $BT_x = \mathbf{u}^T(n_x A)\mathbf{u}$ and $BT_\sigma = \mathbf{u}^T(n_\sigma B)\mathbf{u}$. Consequently, the evolution of the semi-norm is entirely governed by the boundary conditions, implying that Eq. (24) (or Eq. (23)) is energy conserving.

*Remark.* Replacing $\tilde{I}$ in Eq. (24) with the matrix $E = diag(c^{-2}, 1, 1)$ yields the skew-symmetric form of the artificial compressibility method. In this case, Eq. (27) becomes

$$\frac{d\|\mathbf{u}\|_E^2}{dt} + \frac{1}{2}\oint_{\partial\Omega}(BT_x + BT_\sigma)dA = 0, \tag{28}$$

where

$$\|\mathbf{u}\|_E^2 = \int_\Omega J\left[\frac{u^2}{2} + \frac{w^2}{2} + \left(\frac{p}{\rho_o c}\right)^2\right] d\Omega. \tag{29}$$

It should be emphasized that the energy norm (29) is not formally consistent with the definition of physical (thermodynamic) energy. In the artificial compressibility framework, the pressure $p$ is not a thermodynamic state variable but rather a Lagrange multiplier introduced to enforce the incompressibility constraint [17]. Consequently, $p$ is not subject to a positivity-preserving constraint and may take negative values. Indeed, Jung & Giometto [18] demonstrated that, within a family of pseudo-compressibility models, $p$ represents the fluctuating component of the pressure field.

### 2.3. Boundary conditions in the rectangular domain

To examine the boundary contributions, $BT_x$ and $BT_\sigma$ are expanded as follows:

$$\begin{cases} \mathrm{BT}_x = \mathbf{u}^T(n_x A)\mathbf{u} = \left(\dfrac{u^2}{2} + \dfrac{w^2}{2} + \dfrac{p}{\rho_o}\right) J u n_x, \\ \mathrm{BT}_\sigma = \mathbf{u}^T(n_\sigma B)\mathbf{u} = \left[\left(\dfrac{u^2}{2} + \dfrac{w^2}{2} + \dfrac{p}{\rho_o}\right) w^* + z_t \dfrac{p}{\rho_o}\right] n_\sigma, \end{cases} \tag{30}$$

representing the transport of kinetic energy in the $x$- and $\sigma$-directions, as well as the work performed by the pressure associated with the motion of the coordinate system. Along the lateral boundaries, the inflow and outflow of the energy flux play a crucial role. The open boundary conditions are therefore constructed so that the boundary term does not contribute to artificial energy growth, while allowing outgoing energy to leave the domain without spurious reflection. At the top boundary condition ($\sigma = 1$), where $z_t = z_x = 0$, $\mathrm{BT}_\sigma$ in Eq. (30) reduces to

$$\mathrm{BT}_\sigma = \left(\dfrac{u^2}{2} + \dfrac{w^2}{2} + \dfrac{p}{\rho_o}\right) w n_\sigma. \tag{31}$$

Hence, as for the lateral boundaries, the open boundary condition at the top boundary must be constructed to ensure that the energy remains bounded. When a zero-flux condition is imposed at the top boundary, i.e. $w = 0$ at $\sigma = 1$, Eq. (31) further reduces to $\mathrm{BT}_\sigma = 0 n_\sigma$. In contrast, at the terrain layer, i.e., $\sigma = 0$, the impermeable condition can be considered as follows:

$$w^* = w - z_t - z_x u = 0 \text{ at } \sigma = 0 \tag{32}$$

which implies the fluid velocity coincides with the velocity of the bottom surface at the bottom boundary. Thus, at $\sigma = 0$, $\mathrm{BT}_\sigma$ reduces to:

$$\mathrm{BT}_\sigma = w_b \dfrac{p}{\rho_o} n_\sigma. \tag{33}$$

where $w_b = z_t$ represents the vertical velocity of the terrain. Hence, when the bottom boundary moves, work associated with the pressure field is performed. In particular, for $w_b > 0$, the bed does work on the fluid, whereas for $w_b < 0$, the fluid does work on the bed. In the case $w_b = 0$, no pressure work is exchanged across the bottom boundary.

Such boundaries can be implemented by characteristic-based boundary conditions. Firstly, for lateral boundary conditions, we have $BT_x = \mathbf{u}^T A \mathbf{u}$. By diagonalizing $A$, we have $A = P_A \Lambda_A P_A^T$, where $\Lambda_A = diag(\lambda_A^{(1)}, \lambda_A^{(2)}, \lambda_A^{(3)})$,

$$P_A = \begin{pmatrix} \frac{1}{\sqrt{1+\left(\lambda_A^{(1)}\right)^2}} & \frac{1}{\sqrt{1+\left(\lambda_A^{(2)}\right)^2}} & 0 \\ \frac{\lambda_A^{(1)}}{\sqrt{1+\left(\lambda_A^{(1)}\right)^2}} & \frac{\lambda_A^{(2)}}{\sqrt{1+\left(\lambda_A^{(2)}\right)^2}} & 0 \\ 0 & 0 & 1 \end{pmatrix}, \tag{34}$$

with

$$\lambda_A^{(1)} = \frac{u + \sqrt{u^2 + 4}}{2}, \lambda_A^{(2)} = \frac{u - \sqrt{u^2 + 4}}{2}, \lambda_A^{(3)} = u. \tag{35}$$

Therefore, $BT_x = \mathbf{u}^T A \mathbf{u} = W^T \Lambda_A W$ where $W = P_A^T \mathbf{u}$, where the first component of $W$ represents an outgoing characteristic from the interior domain (since $\lambda_A^{(1)} > 0$) and the second component of $W$ corresponds to an incoming characteristic entering the domain from the exterior (since $\lambda_A^{(2)} < 0$). The third component of $W$ may be either incoming or outgoing depending on the sign of $u$ (since $\lambda_A^{(3)} = u$).

It is worth noting that, at the boundary, the number of boundary conditions required for well-posedness equals the number of incoming characteristics, i.e., those associated with negative eigenvalues of the normal flux Jacobian $A_n = n_x A + n_\sigma B$. The boundary operator is maximally semi-bounded if and only if the incoming characteristic variables are prescribed, ensuring that the boundary contribution to the energy estimate satisfies $BT_x + BT_\sigma = \mathbf{u}^T A \mathbf{u} + \mathbf{u}^T B \mathbf{u} \geq 0$, and hence the discrete energy $\|\mathbf{u}\|_I^2$ remains bounded. Therefore, for lateral boundaries, two boundary conditions must be imposed at an inflow case, whereas only one boundary condition is required at an outflow case, corresponding to the number of incoming characteristic variables, so that $BT_x = \mathbf{u}^T A \mathbf{u} = W^T \Lambda_A W \geq 0$.

Similarly, for top and bottom boundaries, $BT_\sigma = \mathbf{u}^T (n_\sigma B) \mathbf{u}$. As $B = P_B \Lambda_B P_B^T$, where $\Lambda_B = diag(\lambda_B^{(1)}, \lambda_B^{(2)}, \lambda_B^{(3)})$,

$$P_B = \begin{pmatrix} \dfrac{-w^* + d}{2N^{(1)}} & \dfrac{-w^* - d}{2N^{(2)}} & 0 \\ \dfrac{-z_x}{N^{(1)}} & \dfrac{-z_x}{N^{(2)}} & \dfrac{1}{\sqrt{1 + z_x^2}} \\ \dfrac{1}{N^{(1)}} & \dfrac{1}{N^{(2)}} & \dfrac{z_x}{\sqrt{1 + z_x^2}} \end{pmatrix}, \tag{36}$$

with $d = \sqrt{w^{*2} + 4z_x^2 + 4}$,

$$N^{(1)} = \sqrt{\left(\dfrac{-w^* + d}{2}\right)^2 + z_x^2 + 1}, \quad N^{(1)} = \sqrt{\left(\dfrac{w^* + d}{2}\right)^2 + z_x^2 + 1}, \tag{37}$$

and

$$\lambda_B^{(1)} = \dfrac{w^* + d}{2J}, \lambda_B^{(2)} = \dfrac{w^* - d}{2J}, \lambda_B^{(3)} = \dfrac{w^*}{J}. \tag{38}$$

Thus, we have $BT_\sigma = \mathbf{u}^T B \mathbf{u} = \widehat{W}^T \Lambda_B \widehat{W}$ where $\widehat{W} = P_B^T \mathbf{u}$. The well-posedness requires prescribing as many boundary conditions as there are negative eigenvalues of $\Lambda_B$. Since $\lambda_B^{(1)} > 0$ and $\lambda_B^{(2)} < 0$, at least one incoming characteristic variable must be specified at the $\sigma$-boundary. At the top boundary ($\sigma = 1$), the total number of required conditions is one or two depending on the sign of $\lambda_B^{(3)}$; in particular, two conditions are needed when $\lambda_B^{(3)} < 0$ (inflow case), whereas only one is needed when $\lambda_B^{(3)} > 0$ (outflow case). At the bottom boundary ($\sigma = 0$), since the normal vector reverses direction, the signs of the eigenvalues are reversed. Consequently, characteristic variables that are outgoing at the top become incoming at the bottom, and vice versa.

## 3. Incompressible Navier-Stokes equations in the sigma-coordinate system

### 3.1. Conservative and Skew-symmetric forms

The incompressible Navier–Stokes equations in the $\sigma$-coordinate system are obtained by replacing the source terms $\tilde{\mathbf{S}}$ in Eqs. (7) and (24) with the following expression:

$$\tilde{\mathbf{S}} = \frac{1}{\rho_o}\begin{pmatrix} 0 \\ \partial_x(J\tau_{11}) - \partial_\sigma(z_x\tau_{11}) + \partial_\sigma\tau_{12} \\ \partial_x(J\tau_{21}) - \partial_\sigma(z_x\tau_{21}) + \partial_\sigma\tau_{22} \end{pmatrix} \tag{39}$$

where $\tau_{ij}$ represents the shear stress tensor. Then, Eq. (26) becomes

$$\frac{d\|\mathbf{u}\|_J^2}{dt} + \frac{1}{2}\oint_{\partial\Omega}(\text{BT}_x + \text{BT}_\sigma)dA = \int_\Omega \mathbf{u}^T\Lambda\tilde{\mathbf{S}}d\Omega, \tag{40}$$

where $\mathbf{u}^T\Lambda\tilde{\mathbf{S}} = u\partial_x(J\tau_{11}) - u\partial_\sigma(z_x\tau_{11}) + u\partial_\sigma\tau_{12} + w\partial_x(J\tau_{21}) - w\partial_\sigma(z_x\tau_{21}) + w\partial_\sigma\tau_{22}$ represents the viscous work associated with the stress divergence. Using Green's theorem, Eq. (40) becomes

$$\frac{d\|\mathbf{u}\|_J^2}{dt} + \frac{1}{2}\oint_{\partial\Omega}(\widehat{\text{BT}}_x + \widehat{\text{BT}}_\sigma)dA = \int_\Omega \varepsilon d\Omega, \tag{41}$$

where

$$\begin{cases} \widehat{\text{BT}}_x = \mathbf{u}^T(n_x A)\mathbf{u} - \rho_o^{-1}(J\tau_{11}u + J\tau_{21}w)n_x, \\ \widehat{\text{BT}}_\sigma = \mathbf{u}^T(n_\sigma B)\mathbf{u} - \rho_o^{-1}(\tau_{12}u - z_x\tau_{11}u - z_x\tau_{21}w + \tau_{22}w)n_\sigma, \\ \varepsilon = \rho_o^{-1}[\tau_{11}(\partial_x(Ju) - \partial_\sigma(z_x u)) + \tau_{12}\partial_\sigma u + \tau_{21}(\partial_x(Jw) - \partial_\sigma(z_x w)) + \tau_{22}\partial_\sigma w]. \end{cases} \tag{42}$$

In addition to the transport of kinetic energy and flow work at the boundary, the energy balance also includes the work associated with shear stresses, while viscous dissipation occurs within the interior of the domain. It is worth noting that, in compressible flows, the viscous dissipation contributes to entropy production in the pressure evolution [17], leading to a cancellation of volumetric dissipation terms in the total energy balance, with only boundary terms remaining (e.g. see [10]). As all variables correspond to thermodynamic quantities in the compressible flows, for an arbitrary control volume, the first law of thermodynamics is satisfied. This implies that the change in internal energy is determined solely by the energy/heat fluxes across the boundary and the work performed at the boundary. By contrast, in incompressible systems, pressure is not a thermodynamic state but rather acts as a Lagrange multiplier enforcing the incompressibility constraint. Consequently, viscous dissipation cannot be absorbed into the pressure evolution process, and therefore remains as an energy sink for viscous flows.

### 3.2 Boundary conditions

To construct the characteristic-based boundary conditions, we need to define the additional vector including the shear stress tensor $\tau_{ij}$. Firstly, for the lateral boundary conditions, the boundary contribution can be re-written as follows:

$$\widehat{BT}_x = \begin{pmatrix} \mathbf{u} \\ \mathbf{v} \end{pmatrix}^T \begin{pmatrix} A & -\frac{1}{2}I_3 \\ -\frac{1}{2}I_3 & 0 \end{pmatrix} \begin{pmatrix} \mathbf{u} \\ \mathbf{v} \end{pmatrix} \text{ with } \mathbf{v} = \begin{pmatrix} 0 \\ \sqrt{J}\tau_{11}/\rho_o \\ \sqrt{J}\tau_{21}/\rho_o \end{pmatrix}, \tag{43}$$

where $I_3 = diag(1,1,1)$. Following [10,11], introducing the non-singular block rotational matrix $R$ defined as

$$R = \begin{pmatrix} I_3 & D \\ \mathbf{0} & I_3 \end{pmatrix} \text{ and } R^{-1} = \begin{pmatrix} I_3 & -D \\ \mathbf{0} & I_3 \end{pmatrix}, \text{s.t.}, R\,R^{-1} = \begin{pmatrix} I_3 & \mathbf{0} \\ \mathbf{0} & I_3 \end{pmatrix}, \tag{44}$$

Eq. (43) can be reformulated as follows:

$$\widehat{BT}_x = \begin{pmatrix} \mathbf{u} \\ \mathbf{v} \end{pmatrix}^T (R\,R^{-1})^T \begin{pmatrix} A & -\frac{1}{2}I_3 \\ -\frac{1}{2}I_3 & 0 \end{pmatrix} R\,R^{-1} \begin{pmatrix} \mathbf{u} \\ \mathbf{v} \end{pmatrix}$$
$$= \left(R^{-1}\begin{pmatrix}\mathbf{u}\\\mathbf{v}\end{pmatrix}\right)^T \begin{pmatrix} A & AD - \frac{1}{2}I_3 \\ D^T A - \frac{1}{2}I_3 & D^T A D - \frac{1}{2}(D + D^T) \end{pmatrix} \left(R^{-1}\begin{pmatrix}\mathbf{u}\\\mathbf{v}\end{pmatrix}\right), \tag{45}$$

To achieve a block-diagonalized form, we can choose $D = 0.5A^{-1}$, then Eq. (45) reduces to

$$\widehat{BT}_x = \begin{pmatrix} \mathbf{u} - \frac{1}{2}A^{-1}\mathbf{v} \\ \mathbf{v} \end{pmatrix}^T \begin{pmatrix} A & 0 \\ 0 & -\frac{1}{4}A^{-1} \end{pmatrix} \begin{pmatrix} \mathbf{u} - \frac{1}{2}A^{-1}\mathbf{v} \\ \mathbf{v} \end{pmatrix}, \tag{46}$$

which leads to

$$\widehat{BT}_x = U_{(1)}^T A U_{(1)} - \frac{1}{4} U_{(2)}^T A^{-1} U_{(2)} = W_{(1)}^T \Lambda_A W_{(1)} - \frac{1}{4} W_{(2)}^T \Lambda_A^{-1} W_{(2)}, \tag{47}$$

where $U_{(1)} = \mathbf{u} - 0.5 A^{-1}\mathbf{v}$, $U_{(2)} = \mathbf{v}$, $W_{(1)} = P_A^T U_{(1)}$, $W_{(2)} = P_A^T U_{(2)}$, and $\Lambda_A^{-1} = diag(1/\lambda_A^{(1)}, 1/\lambda_A^{(2)}, 1/\lambda_A^{(3)})$. Of the six characteristic variables, the three characteristics of $W_{(1)}$ are identical to those of the Euler equations (see Sect 2.2). Moreover, $W_{(2)}$ takes the form:

$$W_{(2)} = \left( \frac{\lambda_A^{(1)} \sqrt{J}\tau_{11}}{\rho_o \sqrt{1 + \left(\lambda_A^{(1)}\right)^2}}, \frac{\lambda_A^{(2)} \sqrt{J}\tau_{11}}{\rho_o \sqrt{1 + \left(\lambda_A^{(2)}\right)^2}}, \frac{\sqrt{J}\tau_{21}}{\rho_o} \right), \tag{48}$$

where the first two components arise from projecting the shear-stress vector onto the first two eigenmodes of $A$. The first component of $W_{(2)}$, associated with the eigenvalue $1/\lambda_A^{(1)}$, is outgoing (i.e., $\lambda_A^{(1)} > 0$), whereas the second component, associated with $1/\lambda_A^{(2)}$, is incoming (i.e., $\lambda_A^{(2)} < 0$). The third component associated with $1/\lambda_A^{(3)}$ depends on the sign of $u$. Since the number of boundary conditions is determined by the number of incoming characteristics, the lateral boundaries require four boundary conditions for inflow cases and two boundary conditions for outflow cases.

For the top and bottom boundaries, boundary contributions can be expressed as

$$\widehat{BT}_\sigma = \begin{pmatrix}\mathbf{u}\\\hat{\mathbf{v}}\end{pmatrix}^T \begin{pmatrix} B & -\frac{1}{2}I_3 \\ -\frac{1}{2}I_3 & 0 \end{pmatrix} \begin{pmatrix}\mathbf{u}\\\hat{\mathbf{v}}\end{pmatrix} \text{ with } \hat{\mathbf{v}} = \begin{pmatrix} 0 \\ \sqrt{J}(\tau_{12} - z_x\tau_{11})/\rho_o \\ \sqrt{J}(\tau_{22} - z_x\tau_{21})/\rho_o \end{pmatrix}. \qquad (49)$$

Through similar procedures with Eqs (43)–(47), Eq. (49) reduces to

$$\widehat{BT}_\sigma = \widehat{U}_{(1)}^T B \widehat{U}_{(1)} - \frac{1}{4} \widehat{U}_{(2)}^T B^{-1} \widehat{U}_{(2)} = \widehat{W}_{(1)}^T \Lambda_B \widehat{W}_{(1)} - \frac{1}{4} \widehat{W}_{(2)}^T \Lambda_B^{-1} \widehat{W}_{(2)}, \qquad (50)$$

where $U_{(1)} = \mathbf{u} - 0.5 B^{-1}\hat{\mathbf{v}}$, $U_{(2)} = \hat{\mathbf{v}}$, $\widehat{W}_{(1)} = P_B^T U_{(1)}$, $\widehat{W}_{(2)} = P_B^T \widehat{U}_{(2)}$, and $\Lambda_B^{-1} = diag(1/\lambda_B^{(1)}, 1/\lambda_B^{(2)}, 1/\lambda_B^{(3)})$. The number of required boundary conditions at the top and bottom boundaries is determined by the number of incoming characteristics, following the same principle as for the lateral boundaries.

*Remark.* The boundary contributions can be re-written as

$$\begin{cases} \widehat{BT}_x = \left[\left(\frac{u^2}{2} + \frac{w^2}{2} + \frac{p}{\rho_o} - \frac{\tau_{11}}{\rho_o}\right)Ju - \frac{\tau_{21}}{\rho_o}Jw\right]n_x, \\ \widehat{BT}_\sigma = \left[\left(\frac{u^2}{2} + \frac{w^2}{2} + \frac{p}{\rho_o}\right)w^* + z_t\frac{p}{\rho_o} - \left(\frac{\tau_{12}}{\rho_o} - z_x\frac{\tau_{11}}{\rho_o}\right)u - \left(\frac{\tau_{22}}{\rho_o} - z_x\frac{\tau_{21}}{\rho_o}\right)w\right]n_\sigma. \end{cases} \qquad (51)$$

Compared to Eq. (30), Eq. (51) includes an additional boundary work term associated with the deviatoric (viscous) stresses. At the top boundary, i.e., $\sigma = 1$, where $z_t = z_x = 0$,

$$\widehat{BT}_\sigma = \left[\left(\frac{u^2}{2} + \frac{w^2}{2} + \frac{p}{\rho_o} - \frac{\tau_{22}}{\rho_o}\right)w - \frac{\tau_{12}}{\rho_o}u\right]n_\sigma. \qquad (52)$$

For the zero-flux condition, Eq. (53) reduces to:

$$\widehat{BT}_\sigma = -\left[\left(\frac{\tau_{12}}{\rho_o} - z_x \frac{\tau_{11}}{\rho_o}\right) u\right] n_\sigma. \tag{53}$$

At the terrain layer ($\sigma = 0$), if the impermeable condition $w^* = 0$ is imposed, Eq. (51) reduces to

$$\widehat{BT}_\sigma = \left[z_t \frac{p}{\rho_o} - \left(\frac{\tau_{12}}{\rho_o} - z_x \frac{\tau_{11}}{\rho_o}\right) u - \left(\frac{\tau_{22}}{\rho_o} - z_x \frac{\tau_{21}}{\rho_o}\right)(z_t + z_x u)\right] n_\sigma. \tag{54}$$

If, in addition, the no-slip condition is imposed along the bottom boundary, Eq. (54) further reduces to

$$\widehat{BT}_\sigma = \left[\frac{p}{\rho_o} - \left(\frac{\tau_{22}}{\rho_o} - z_x \frac{\tau_{21}}{\rho_o}\right)\right] z_t n_\sigma, \tag{55}$$

which implies that the sign of the bottom boundary work is governed by the terrain velocity $z_t$; depending on its direction, the moving bottom may either inject energy into or remove energy from the flow.

## 4. Summary and conclusions

In this study, conservative and skew-symmetric formulations have been derived for incompressible flows over movable irregular terrain. First, a conservative formulation of the general conservation laws has been derived in the $\sigma$-coordinate system. Building upon this structure, a pseudo-compressibility-based approach has been introduced, providing the well-known advantages of conservative formulations (e.g., shock-capturing capability) to be fully utilized. For the skew-symmetric formulation, it has been shown that, in the inviscid case (Euler equations), a semi-energy norm rate is entirely determined by the surface-interface contribution, thereby yielding an energy-conserving property. In the viscous case (Navier–Stokes equations), the formulation becomes energy-bounded due to the presence of viscous dissipation. It is further discussed that this dissipation term arises from the modified role of pressure: under the incompressibility constraint, pressure ceases to be a thermodynamic variable and instead functions as a Lagrange multiplier. Boundary conditions have also been briefly investigated, and their physical implications have been examined within the energy analysis framework.

When choosing between the two formulations, one has to keep in mind their overall modeling objective. When modeling incompressible flows over irregular topography using pseudo-compressibility–based approaches, the conservative formulation is recommended, as it allows direct application of Riemann-solver theory and facilitates the construction of robust numerical schemes. In contrast, for simulations targeting the accurate resolution of turbulent structures in smooth flows, the skew-symmetric formulation is advantageous due to its energy-bounded property.

# Acknowledgement

This work is supported by the Office of Naval Research, Young Investigator Program (YIP), grant no. N000142412604.

# References

[1] Thompson, J. F., Warsi, Z. U., & Mastin, C. W. (1985). Numerical grid generation: foundations and applications. Elsevier North-Holland, Inc..

[2] Mittal, R., & Iaccarino, G. (2005). Immersed boundary methods. *Annu. Rev. Fluid Mech.*, *37*(1), 239-261.

[3] Gal-Chen, T., & Somerville, R. C. (1975). On the use of a coordinate transformation for the solution of the Navier-Stokes equations. Journal of Computational Physics, 1, 209-228.

[4] Yang, D., & Shen, L. (2011). Simulation of viscous flows with undulatory boundaries. Part I: Basic solver. Journal of Computational Physics, 230, 5488-5509.

[5] Klemp, J. B. (2011). A terrain-following coordinate with smoothed coordinate surfaces. Monthly Weather Review, 139, 2163-2169.

[6] Schär, C., Leuenberger, D., Fuhrer, O., Lüthi, D., & Girard, C. (2002). A new terrain-following vertical coordinate formulation for atmospheric prediction models. Monthly Weather Review, 130, 2459-2480.

[7] Chassignet, E. P., Hurlburt, H. E., Smedstad, O. M., Halliwell, G. R., Hogan, P. J., Wallcraft, A. J., & Bleck, R. (2006). Ocean prediction with the hybrid coordinate ocean model (HYCOM). In Ocean weather forecasting: an integrated view of oceanography (pp. 413-426). Dordrecht: Springer Netherlands.

[8] Nordström, J. (2022). A skew-symmetric energy and entropy stable formulation of the compressible Euler equations. Journal of Computational Physics, 470, 111573.

[9] Nordström, J. (2022). Nonlinear and linearised primal and dual initial boundary value problems: When are they bounded? How are they connected?. Journal of Computational Physics, 455, 111001.

[10] Nordström, J. (2024). A skew-symmetric energy stable almost dissipation free formulation of the compressible Navier-Stokes equations. Journal of Computational Physics, 512, 113145.

[11] Nordström, J., & Malan, A. G. (2025). An energy stable incompressible multi-phase flow formulation. Journal of Computational Physics, 523, 113685.

[12] Gustafsson, B., Kreiss, H. O., & Oliger, J. (1995). *Time dependent problems and difference methods* (Vol. 24). John Wiley & Sons.

[13] Jung, J., Schmid, M., Fish, J., Weng, E., & Giometto, M. (2025). Path-conservative well-balanced high-order finite-volume solver for the volume-averaged Navier–Stokes equations with discontinuous porosity. Journal of Computational Physics, 533, 113978.

[14] Han, E. E., Hantke, M., & Warnecke, G. (2012). Exact Riemann solutions to compressible Euler equations in ducts with discontinuous cross-section. Journal of Hyperbolic Differential Equations, 9, 403-449.

[15] Toro, E. F. (2013). Riemann solvers and numerical methods for fluid dynamics: a practical introduction. Springer Science & Business Media.

[16] Chorin, A. J. (1967). A numerical method for solving incompressible viscous flow problems. Journal of Computational Physics, 2, 12-26.

[17] Jung, J., & Giometto, M. (2026). A general formulation for pseudo-compressibility methods for the simulation of fluid flow. *Available at SSRN 6137887*.

[18] Jung, J., & Giometto, M. (2026). Consideration on the entropically-damped artificial-compressibility method. *Available at SSRN 6560008*

# Appendix. Derivation of Eq. (23)

Firstly, the continuity equation can be re-formulated as follows

$$\begin{aligned}\frac{\partial (Ju)}{\partial x} + \frac{\partial (w - z_x u)}{\partial \sigma} &= \sqrt{J}\left(\frac{\partial U}{\partial x} + \frac{1}{J}\frac{\partial W}{\partial \sigma} - \frac{z_x}{J}\frac{\partial U}{\partial \sigma} - \frac{U}{2J}\frac{\partial z_x}{\partial \sigma}\right) \\ &= \sqrt{J}\left(\frac{\partial U}{\partial x} + \frac{\partial}{\partial \sigma}\left(\frac{W}{J}\right) - \frac{\partial}{\partial \sigma}\left(\frac{z_x U}{J}\right) + \frac{U}{2J}\frac{\partial z_x}{\partial \sigma}\right) \\ &= \sqrt{J}\left(\frac{1}{2}\left[\frac{\partial 1U}{\partial x} + 1\frac{\partial U}{\partial x}\right] + \frac{1}{2}\left[\frac{\partial}{\partial \sigma}\left(\frac{W}{J}\right) + \frac{1}{J}\frac{\partial W}{\partial \sigma}\right] - \frac{1}{2}\left[\frac{\partial}{\partial \sigma}\left(\frac{z_x U}{J}\right) + \frac{z_x}{J}\frac{\partial U}{\partial \sigma}\right]\right).\end{aligned} \quad (56)$$

Secondly, the horizontal momentum equation is recast. Each term in the $x$-directional momentum balance equation in Eq. (20) can be re-written as:

(i)
$$\begin{aligned}\frac{\partial (Ju)}{\partial t} &= \sqrt{J}\frac{\partial U}{\partial t} + U\frac{J_t}{2\sqrt{J}} \\ &= \sqrt{J}\frac{\partial U}{\partial t} + \frac{u}{2}\frac{\partial J}{\partial t},\end{aligned} \quad (57)$$

(ii)
$$\begin{aligned}\frac{1}{2}\left[\frac{\partial (Juu)}{\partial x} + Ju\frac{\partial u}{\partial x}\right] &= \frac{1}{2}\left[2U\frac{\partial U}{\partial x} + \left(U\frac{\partial U}{\partial x} - \frac{U^2}{2J}J_x\right)\right] \\ &= \frac{3}{2}UU_x - \frac{U^2}{4J}J_x \\ &= \frac{\sqrt{J}}{2}\left(3\frac{U}{\sqrt{J}}U_x - \frac{U^2}{2J^{\frac{3}{2}}}J_x\right) \\ &= \frac{\sqrt{J}}{2}(u_x U + 2uU_x) \\ &= \frac{\sqrt{J}}{2}[(Uu)_x + uU_x],\end{aligned} \quad (58)$$

(iii)
$$\begin{aligned}\frac{1}{2}\left[\frac{\partial (uw^*)}{\partial \sigma} + w^*\frac{\partial u}{\partial \sigma}\right] &= \frac{U}{2\sqrt{J}}\frac{\partial (w^*)}{\partial \sigma} + \frac{w^*}{\sqrt{J}}\frac{\partial U}{\partial \sigma} - w^*U\frac{J_\sigma}{2J^{3/2}} \\ &= \frac{\sqrt{J}}{2}\left(\frac{U}{J}\frac{\partial w^*}{\partial \sigma} + 2\frac{w^*}{J}U_\sigma - w^*U\frac{J_\sigma}{J^2}\right) \\ &= \frac{\sqrt{J}}{2}\left(U\frac{\partial}{\partial \sigma}\left(\frac{w^*}{J}\right) + 2\frac{w^*}{J}U_\sigma\right) \\ &= \frac{\sqrt{J}}{2}\left[\frac{\partial}{\partial \sigma}\left(\frac{w^*}{J}U\right) + \frac{w^*}{J}U_\sigma\right],\end{aligned} \quad (59)$$

(iv)
$$\begin{aligned}\frac{\partial}{\partial x}\left(J\frac{p}{\rho_o}\right) - \frac{\partial}{\partial \sigma}\left(z_x\frac{p}{\rho_o}\right) &= \frac{\partial}{\partial x}(\sqrt{J}P) - \frac{\partial}{\partial \sigma}\left(\frac{z_x}{\sqrt{J}}P\right) \\ &= \sqrt{J}\left(\frac{\partial P}{\partial x} - \frac{z_x}{J}\frac{\partial P}{\partial \sigma} - \frac{P}{2J}\frac{\partial J}{\partial x}\right)\end{aligned} \quad (60)$$

$$= \sqrt{J}\left(\frac{\partial P}{\partial x} - \frac{\partial}{\partial \sigma}\left(\frac{z_x}{J}P\right) + \frac{P}{2J}\frac{\partial J}{\partial x}\right)$$

$$= \frac{\sqrt{J}}{2}\left(\frac{\partial(1P)}{\partial x} + 1\frac{\partial P}{\partial x}\right) - \frac{\sqrt{J}}{2}\left(\frac{\partial}{\partial \sigma}\left(\frac{z_x}{J}P\right) + \frac{z_x}{J}\frac{\partial P}{\partial \sigma}\right).$$

Note that, in Eq. (60), the final expression is obtained by averaging the second and third lines. Introducing Eqs. (57) -(60) into the $x$-directional momentum balance equation in Eq. (20) yields

$$\frac{\partial U}{\partial t} + \frac{1}{2}[(Uu)_x + uU_x] + \frac{1}{2}\left[\frac{\partial}{\partial \sigma}\left(\frac{w^*}{J}U\right) + \frac{w^*}{J}U_\sigma\right] + \frac{1}{2}\left[\frac{\partial(1P)}{\partial x} + 1\frac{\partial P}{\partial x}\right] - \frac{1}{2}\left[\frac{\partial}{\partial \sigma}\left(\frac{z_x}{J}P\right) + \frac{z_x}{J}\frac{\partial P}{\partial \sigma}\right] = 0. \quad (61)$$

Following a similar procedure, the $\sigma$-directional component of momentum balance equations can be written as

$$\frac{\partial W}{\partial t} + \frac{1}{2}[(Wu)_x + uW_x] + \frac{1}{2}\left[\frac{\partial}{\partial \sigma}\left(\frac{w^*}{J}W\right) + \frac{w^*}{J}W_\sigma\right] + \frac{1}{2}\left[\frac{\partial}{\partial \sigma}\left(\frac{1}{J}P\right) + \frac{1}{J}\frac{\partial P}{\partial \sigma}\right] = 0. \quad (62)$$